\documentclass[10pt]{article}
\usepackage{amssymb,dsfont,amsmath,graphics,colortbl,epsfig}

\def\biglb{\big[\hspace*{-.7mm}\big[}
\def\bigrb{\big]\hspace*{-.7mm}\big]}
\def\Biglb{\Big[\hspace*{-1.4mm}\Big[}
\def\Bigrb{\Big]\hspace*{-1.3mm}\Big]}

\def\openone{\leavevmode\hbox{\small1\kern-3.3pt\normalsize1}}
\def\res{\mathop{\mbox{Res}\,}\limits}

\def\Cup{\mathop{\cup}\limits}

\def\tr{\mathrm{tr\,}}

\def\im{\mathrm{Im\,}}
\def\det{\mathrm{det\,}}
\def\ad{\mathrm{ad\,}_{\sigma_{3 }}}

\def\diag{\mbox{diag\,}}

\def\bbbr{{\Bbb R}}
\begin{document}
\arraycolsep=2pt

\numberwithin{equation}{section}
\allowdisplaybreaks
\bibliographystyle{plain}

\title{On the multi-component nonlinear Schr\"odinger equation with
constant boundary conditions}

\author{V. A. Atanasov,  V. S. Gerdjikov\\
\it Institute of Nuclear Research and Nuclear Energy, \\[-1.mm]
\it Bulgarian Academy of Sciences, Sofia 1784, Bulgaria \\[-1.mm]
e-mail: victor@inrne.bas.bg\\
e-mail: gerjikov@inrne.bas.bg}
\date{ }

\maketitle

\begin{abstract}
The multi-component nonlinear Schr\"odinger equation related to
${\bf C.I}\simeq Sp(2p)/U(p)$  and  ${\bf D.III}\simeq
SO(2p)/U(p)$-type symmetric spaces with non-vanishing boundary
conditions is solvable with the inverse scattering method (ISM).
As Lax operator $L$ we use the generalized Zakharov-Shabat
operator. We show that the ISM for the Lax operator $L(x,
\lambda)$ is a nonlinear analog of the Fourier-transform
method. As appropriate generalizations of the usual
Fourier-exponential functions we use the so-called ''squared
solutions'', which are constructed in terms of the fundamental
analytic solutions (FAS) $\chi^{\pm}(x, \lambda)$ of $L(x,
\lambda)$ and the Cartan-Weyl basis of the Lie algebra, relevant to the
symmetric space. We derive the completeness relation
for the "squared solutions" which turns out to provide spectral
decomposition of the recursion (generating) operators
$\Lambda_{\pm}$, a natural generalizations of
$\frac{1}{i}\frac{d}{dx}$ in the case of nonlinear evolution
equations (NLEE).
\end{abstract}

\section{Introduction}

The integrability of the scalar nonlinear Schr\"odinger  equation
(NLS) with vanishing boundary conditions (v.b.c.):
\begin{equation}\label{1}
    iq_{t}+q_{xx}+2|q(x,t)|^2 q(x,t)=0
\end{equation}
was discovered by Zakharov and Shabat in their pioneer
work \cite{ZSh1}. Soon after  \cite{ZSh2} Zakharov and Shabat proved
the integrability and the physical importance of the NLS with
constant boundary conditions (c.b.c.):
\begin{equation}\label{2}
iq_{t}+2q_{xx}-2(|q(x,t)|^2-\rho^2)q(x,t)=0,\qquad
\lim_{x\to\pm\infty}q(x,t)=q_{\pm},
\end{equation}
where the asymptotic values $q_{\pm}$ satisfy $|q_{\pm}|^2=\rho^2$. Notice
the sign difference in the cubic nonlinearity as well as the additional
term with the chemical potential $\rho$.

Both versions of NLS equation served as models on which generalizations
were made. The simplest non-trivial multicomponent generalization of NLS
is the vector NLS known as the Manakov model \cite{Ma*76aR}:
\begin{equation}\label{3}
     i\overrightarrow{q}_{t}+\overrightarrow{q}_{xx}+2(\overrightarrow{q}^
     {\dagger} \overrightarrow{q}(x,t)) \overrightarrow{q}(x,t)=0,
\end{equation}
where $\overrightarrow{q}(x,t)$ is an $n$-component complex-valued
vector vanishing fast enough for $x\to\pm\infty$. The c.b.c.
version of vector NLS
\begin{equation}\label{4}
i\overrightarrow{q}_{t}+\overrightarrow{q}_{xx}-2\left((\overrightarrow{q}^
{\dagger}
\overrightarrow{q}(x,t))-\rho^2\right) \overrightarrow{q}(x,t)=0,
\end{equation}
where
$\lim_{x\to\pm\infty}\overrightarrow{q}(x,t)=\overrightarrow{q}_{\pm}$
and $\overrightarrow{q}_{-}=U_{0}\overrightarrow{q}_{+}$  where
$U_0$ is constant unitary matrix also finds applications. Here
$\rho^2=\overrightarrow{q}^{\dagger}_{\pm}\overrightarrow{q}_{\pm}$.

Equations (\ref{1}) and (\ref{3}) are particular cases of matrix
NLS which is obtained from the system:
\begin{eqnarray}\label{matrix_NLS}
i\textbf{q}_{t}+\textbf{q}_{xx}+2\textbf{q}\textbf{r}\textbf{q}(x,t)=0&,&
-i\textbf{r}_{t}+\textbf{r}_{xx}+2\textbf{r}\textbf{q}\textbf{r}(x,t)=0,
\end{eqnarray}
after imposing appropriate involution (reduction) compatible with
the evolution of (\ref{matrix_NLS}). Here $\textbf{q}$ and
$\textbf{r}$ are $n\times m$ matrix-valued functions of $x$ and
$t$. One such involution is:
\begin{equation}\label{involution}
\textbf{r}=B_{-}\textbf{q}^{\dagger}\,B^{-1}_{+},\qquad B_{\pm}=
\diag(\epsilon^{\pm}_{1},...,\epsilon^{\pm}_{m}),\quad(\epsilon^{\pm}_{1})^2=1,
\end{equation}
and the corresponding MNLS acquires the form:
\begin{equation}\label{NLS}
i\textbf{q}_{t}+\textbf{q}_{xx}+2\textbf{q}B_{-}\textbf{q}^{\dagger}\,B^{-1}_{+}\textbf{q}=0,
\end{equation}
For $n=m=1$ and $r=q^{*}$ the system goes into the scalar NLS
(\ref{1}); for $m=1$ and $n>1$ and with appropriate choice of
involution (\ref{involution}) the system is transformed into the
Manakov model (\ref{3}). All these versions are solvable with the
ISM. The ISM is applicable to nonlinear evolution equations (NLEE)
if they can be represented as compatibility condition of two
linear problems \cite{FaTa,ZMNP,AKNS,AbwSeg}:
\begin{equation}\label{commutator}
\left[L(\lambda), M(\lambda)\right]=0,
\end{equation}
which holds identically with respect to the spectral parameter
$\lambda$.

The two linear operators $L(\lambda)$ and $M(\lambda)$ in the
Zakharov-Shabat system (Z-Sh) for the MNLS on symmetric spaces
associated with the simple Lie algebra $\mathfrak{g}\simeq {\bf
C_{r}}$ and $\mathfrak{g}\simeq {\bf D_{r}}$ with (v.b.c.) are:
\begin{eqnarray}\label{L}
L\psi&=&\left(i\frac {\partial}{\partial x} + Q(x,t)-\lambda\,
\sigma_{3}\right)\psi(x,t,\lambda)=0,\\ \label{M}
M\psi&=&\left(i\frac {\partial}{\partial t}+V_{2}(x,t)+\lambda
V_{1}(x,t)-2\lambda^2 \sigma_{3}\right)\psi(x,t,\lambda)=0,\\
\nonumber\\ \label{Q}
Q(x,t)&=&\left(%
\begin{array}{cc} 0 & \textbf{q}(x,t) \\ \textbf{r}(x,t) & 0 \\
\end{array} \right),\quad \sigma_{3}=\left(\begin{array}{cc}
  \openone & 0 \\   0 & -\openone \\ \end{array} \right),
\end{eqnarray}
where $Q(x,t)$ and $\sigma_{3}$ are $2r\times 2r$ matrices with
compatible block structure. Here
\begin{eqnarray}\label{V2v.b.c.}
V_{1}(x,t)=2Q(x,t), && V_{2}(x,t)=[\ad^{-1}Q,
Q\,]+2i\ad^{-1}Q_{x}(x,t)
\end{eqnarray}
and $\ad^{-1}$ is the inverse of the adjoint action $\ad$ with
respect to the element $\sigma_{3}$: $\ad Y = [\sigma_{3},Y]$.

An  effective tool to obtain new versions of MNLS is the reduction
group introduced by Mikhailov \cite{Mik}. It allows one to impose
algebraic constraints on the potential $Q(x,t)$ which are
automatically compatible with the evolution. For example, the
involution (\ref{involution}), which leads to MNLS with v.b.c.
(\ref{NLS}) is known as $\mathbb{Z}_2$-reduction and can be
written as \cite{GGK}:
\begin{equation}\label{reduction1}
    BU^{\dagger}(x,t,\lambda^{*})B^{-1}=U(x,t,\lambda)
\end{equation}
where $B$ is an automorphism of $\mathfrak{g}$ matrix such that
$B^2=\openone,\;[\sigma_3 ,B]=0,$ and
\begin{equation}
    U(x,t,\lambda)=Q(x,t)-\lambda\, \sigma_{3}.
\end{equation}

Below we analyze the multi-component nonlinear Sch\"odinger
equation (MNLS):
\begin{equation}\label{MNLS}
i\textbf{q}_{t}+\textbf{q}_{xx}-2\textbf{q}\textbf{q}^{\dagger}\,\textbf{q}
+\textbf{q}\overline{\mu}+\mu\,\textbf{q}=0,
\end{equation}
with {\it constant} boundary conditions (c.b.c.) at $x \to \pm
\infty$:
\begin{equation}
\lim_{x\to\pm\infty}\textbf{q}(x,t)=\textbf{q}_{\pm}\,,\qquad \mu=
\textbf{q}_{+}\textbf{q}^{\dagger}_{+}=\textbf{q}_{-}\textbf{q}^{\dagger}_{-}\,,\qquad
\overline{\mu}=\textbf{q}^{\dagger}_{+}\textbf{q}_{+}=\textbf{q}^{\dagger}_{-}\textbf{q}_{-}\,,
\end{equation}
where $\textbf{q}(x, t)$ is  $n\times r$ matrix-valued function,
related to ${\bf A.III}$, ${\bf C.I}$ or ${\bf D.III}$-type
symmetric spaces. The case $n \neq r$ can be related to ${\bf
A.III}$ symmetric spaces only and has been solved with the ISM in
\cite{LOMI}.

Therefore we concentrate on the MNLS (\ref{MNLS}) related to ${\bf
C.I}$ or ${\bf D.III}$-type symmetric spaces, which means in
particular that $n=r$. Its Lax pair is obtained from
(\ref{L})--(\ref{V2v.b.c.}) by replacing $V_2(x,t)$ with:
\begin{equation}\label{V2}
V_{2}(x,t)=[\ad^{-1}Q,Q]+2i\ad^{-1}Q_{x}(x,t)-\sigma_3
Q^{2}_{\pm}.
\end{equation}
Here we have also imposed the additional condition
$Q_{+}^2=Q_{-}^2$. It ensures that the two asymptotic Lax
operators $L_{\pm}=i\frac{d}{dx}+Q_{\pm}-\lambda \sigma_3 $ have
the same spectrum. It also ensures that the potentials
$V_{1}(x,t)$ and $V_{2}(x,t)$ in the second operator $M(\lambda)$
vanish for $x\to\pm\infty$. As a result the solutions of the MNLS
(\ref{MNLS}) $\textbf{q}(x,t)$ do not undergo strong oscillations
with respect to time, see \cite{LOMI,VSG_BAN}.

Lax operators of the form (\ref{L}) can be associated with each of
the symmetric spaces listed below (for the definition see
\cite{Hel} and the Appendix). They are defined by specifying the
simple Lie algebra $\mathfrak{g}$, having typical representation
in $2r\times 2r$ matrices and the Cartan subalgebra element
$\sigma_3$:
\begin{itemize}
    \item ${\bf C.I}$:  $\mathfrak{g} \simeq {\bf C}_{r}\simeq
    sp(2r)$, $\sigma_3=H_{\vec{a}}$, where the vector $\vec{a}$ in
    the root space $\mathbb{E}^{r}$ dual to $\sigma_3$ is given by
    $\vec{a}=\sum_{k=1}^{r}e_k$
    \item ${\bf D.III}$ :  $\mathfrak{g} \simeq {\bf D}_{r}\simeq
    so(2r)$, $\sigma_3=H_{\vec{a}}$, where the vector $\vec{a}$ in
    the root space $\mathbb{E}^{r}$ dual to $\sigma_3$ is given by
    $\vec{a}=\sum_{k=1}^{r}e_k$
\end{itemize}
Here the orthonormal vectors $e_k$ span the root space
$\mathbb{E}^{r}$ of both types of algebras. The element
$\sigma_{3}$ belongs to the Cartan subalgebra $\mathfrak{h}$ and
is dual to $\vec{a}$. Using $\sigma_{3}$ we can split the set of
positive roots into two two subsets
$\triangle^{+}=\triangle_{0}^{+}\cup\triangle_{1}^{+}$. These
sets, for the algebras that we are working with, are composed of
the following roots:
\begin{equation}\label{roots}
\triangle_{0}^{+}\equiv\{e_{i}-e_{j}\,,\;1\leq i<j\leq
r\}\,,\qquad \triangle_{1}^{+}\equiv\{\,2e_{i},\,
e_{i}+e_{j}\,,\;1\leq i<j\leq r\}\end{equation} for
$\mathfrak{g}\simeq sp(2r)$ and
\begin{equation}
\triangle_{0}^{+}\equiv\{e_{i}-e_{j}\,,\;1\leq i<j\leq
r\}\,,\qquad \triangle_{1}^{+}\equiv\{ e_{i}+e_{j}\,,\;1\leq
i<j\leq r\}\end{equation} for $\mathfrak{g}\simeq so(2r)$.

The root vectors of the algebra are denoted by $E_{\alpha}$ where
$\alpha$ is the corresponding root.

Let us introduce a projector $P_{\sigma_{3}}=\ad^{-1}\ad$ onto the
co-adjoint orbit $O_{\sigma_3 }$ of the element $\sigma_3$. Here
the inverse of the adjoint action is $\ad^{-1}Y=\frac12 \sigma_{3}
Y$. The generic element of $X\in O_{\sigma_3 }$ is the one that
satisfies the relation $X=P_{\sigma_{3}}\,X$. Obviously the
potential of the Z-Sh system $Q(x,t)$ and its variation $\delta
Q(x,t)$ belong to $O_{\sigma_3 }$.

This paper extends the results of \cite{LOMI,VSG_BAN}. In Section
2 we focus on the solutions of the direct scattering problem for
the case of Lax operator describing MNLS with c.b.c.. In Section 3
we derive the completeness relation for the "squared solutions" of
the Lax operator generalizing the results of \cite{DJ,KonVek}.
Here we prove that the ISM is equivalent to a generalized Fourier
transform also for Lax operators with c.b.c.  Thus we have shown
that the nonlinear evolution of equation (\ref{MNLS}) transforms
into linear one in terms of the scattering data of $L$.

\section{Solutions of the Lax operator $L$}

The spectrum of the asymptotic operators $L_{\pm}$ is purely
continuous and is determined by the eigenvalues of $Q_{\pm}$ which
generically may be arbitrary complex numbers. However, here we
consider only the case when $L$ becomes self-adjoint. As a result
its potential $Q(x, t)$ acquires the form:
\begin{equation}\label{Qself}
Q(x, t)=-Q^{\dagger}(x,t)\qquad
Q(x,t)=\left(%
            \begin{array}{cc}
                            0 & \textbf{q}(x,t) \\
                    -\textbf{q}^{\dagger}(x,t) & 0 \\
            \end{array}%
        \right)
\end{equation}

For simplicity reasons we will consider only the case when all of
the eigenvalues of the asymptotic matrices $Q_{\pm}$ are real and
equal:
\begin{equation}\label{m}
    m_{1}=m_{2}=...=m_{r}=m\neq0\qquad m\in\bbbr
\end{equation}
As a result we have the following condition on the eigenvalues of
the asymptotic matrices \cite{LOMI}:
$\textbf{q}_{\pm}\textbf{q}^{\dagger}_{\pm}(x,t)=m^2 \openone$ and
the correspondence with the isotropic problem is obvious:
$\mu=\overline{\mu}=m^2 \openone$.

The requirement that the potentials of the Z-Sh system belong to
$\mathfrak{g}$ can be formulated as a reduction condition \cite{Mik,FK}:
\begin{eqnarray}\label{reduct}
S^{-1}_{0}U^{t}(x,t,\lambda)S_{0}=-U(x,t,\lambda),\quad
S^{-1}_{0}V^{t}(x,t,\lambda)S_{0}=-V(x,t,\lambda), &&
S^{-1}_{0}\sigma_{3}S_{0} =-\sigma_{3},
\end{eqnarray}
which has trivial action on $\lambda$. The matrix $S_{0}$ is the
one which realizes the definition of the algebras ${\bf
C}_{r}\simeq sp(2r)$ or ${\bf D}_{r}\simeq so(2r)$ in the typical
representation  \cite{FK,Hel} . In what follows we will define the
Lie algebra $\mathfrak{g}$ by:
\begin{equation}\label{algebra}
    \mathfrak{g}\equiv \left\{X :
X+S^{-1}_{0}X^{t}S_{0}=0\right\},
\end{equation}
where
\begin{equation*}
    S_{0}=\sum_{s=1}^{r}(-1)^{s+1}(E_{s\overline{s}}-E_{\,\overline{s}s})
\end{equation*}for $\mathfrak{g}\simeq sp(2r)$ and
\begin{equation*}
S_{0}=\sum_{s=1}^{r}(-1)^{s+1}(E_{s\overline{s}}+E_{\,\overline{s}s})
\end{equation*} for $\mathfrak{g}\simeq so(2r)$. Here
$\overline{s}=2r-s+1$ and $E_{ks}$ are $2r\times 2r$ matrices,
defined by $(E_{ks})_{ij}=\delta_{ki}\delta_{sj}$. Note that
$S_{0}^2=\epsilon_{0}\openone$, where $\epsilon_{0}=-1$ for
$sp(2r)$ and $\epsilon_{0}=1$ for $so(2r)$.

Such reduction (\ref{reduct}) imposes restrictions only on the
coefficients of $Q(x,t)$ such that for ${\bf C}_{r}\simeq sp(2r)$
we can put:
\begin{equation}
Q(x,t)=\sum_{i<j}\left(q_{ij}E_{e_{i}+e_{j}}-q^{*}_{j\,i}E_{-e_{i}-e_{j}}\right)+
\sum_{i=1}^{r}\left(q_{i}E_{2e_{i}}-q^{*}_{i}E_{-2e_{i}}\right),
\end{equation}
while in the  ${\bf D}_{r}\simeq so(2r)$-case we have:
\begin{equation}
Q(x,t)=\sum_{i<j}\left(q_{ij}E_{e_{i}+e_{j}}-q^{*}_{j\,i}E_{-e_{i}-e_{j}}\right),
\end{equation}
where $^{*}$ means complex conjugation. The definitions of the
root vectors $E_{\alpha}$ can be found in the Appendix. In the
typical representations of ${\bf C}_{r}$ and ${\bf D}_{r}$ these
choices for $Q(x,t)$ have always the block structure shown in
(\ref{Qself}). In the case of $\mathfrak{g}\simeq sp(4)$ the block
$\textbf{q}$ is parametrized by three functions:
\begin{equation}\label{q_sp4}
\textbf{q}(x,t)=\left(%
\begin{array}{cc}
  q_{12} & \sqrt{2} q_{1} \\
  \sqrt{2} q_{2} & -q_{12} \\
\end{array}%
\right).
\end{equation}

The corresponding sets of MNLS for these choices of $Q(x,t)$ and
v.b.c. were first derived in  \cite{FK}. For c.b.c. with
$\textbf{r}=-\textbf{q}^{\dagger}$ MNLS take the form (\ref{MNLS})
with the additional linear in $\textbf{q}$ terms ensuring regular
behavior of the solutions for $t\to\pm\infty$.

Let us outline the construction of the fundamental analytic
solutions (FAS). In the particular case that we are considering -
the isotropic problem - the Jost solutions are defined as
fundamental solutions with fixed asymptotic for $x\to\pm\infty$:
\begin{eqnarray}\label{Jost}
\lim_{x\to\infty}\psi(x, \lambda)e^{ij(\lambda)\sigma_3
x}=\psi_{0}(\lambda)&& \lim_{x\to-\infty}\phi(x,
\lambda)e^{ij(\lambda)\sigma_3x}=\phi_{0}(\lambda),
\end{eqnarray}
where $2r\times 2r$ matrices $\psi_{0}(\lambda)$ and
$\phi_{0}(\lambda)$ take value in the corresponding group
$\mathcal{G}$ and diagonalize the potential of the Lax operator
$L$
\begin{eqnarray}
\left(Q_{+}-\lambda\sigma_3\right)\psi_{0}(\lambda)=-\psi_{0}(\lambda)
j(\lambda)\sigma_{3},\qquad  \left(Q_{-}-\lambda\sigma_3\right)
\phi_{0}(\lambda)=-\phi_{0}(\lambda)j(\lambda)\sigma_{3},
\end{eqnarray}
where $j(\lambda)=\sqrt{\lambda^2-m^2}$. They have the block structure:
\begin{equation}\label{}
\psi_{0}(\lambda)=\left(%
\begin{array}{cc}
  \underline{A} & S_{1}\,\underline{B} \\
  \underline{B}\,S_{1} & \underline{A} \\
\end{array}%
\right), \qquad \phi_{0}(\lambda)=V_0 \,\left(%
\begin{array}{cc}
  \underline{A} & S_{1}\,\underline{B} \\
  \underline{B}\,S_{1} & \underline{A} \\
\end{array}%
\right).
\end{equation}
The $r \times r$ matrices \underline{A}, \underline{B} and $S_{1}$
are given by:
\begin{eqnarray}\label{AB}
\underline{A}_{\,kl}=\delta_{kl}
\sqrt{\frac{\lambda+j(\lambda)}{2j(\lambda)}},\qquad
\underline{B}_{\,kl}=\delta_{kl}
\sqrt{\frac{\lambda-j(\lambda)}{2j(\lambda)}}, \qquad
S_{1}=\sum_{s=1}^{r}(-1)^{s+1}e_{\,s\,,r-s+1}
\end{eqnarray}
where $e_{p\,q}$ are $r \times r$ matrices such that
$(e_{p\,q})_{ij}=\delta_{ip}\,\delta_{jq}$ and the phase factor
$V_0$ is $2r \times 2r$ diagonal and unitary matrix.

The two Jost solutions are  fundamental solutions and must be
linearly dependent. This means that there exists a matrix
$T(t,\lambda)$, called scattering matrix, which connects them and
has an appropriate block structure.
\begin{equation}\label{T}
 T(t, \lambda)=\psi^{-1}(x, t, \lambda)\,\phi(x, t, \lambda)
\end{equation}

The spectral parameter $\lambda$ takes values in two-sheeted
Riemannian surface $\mathcal{S}$:
$$\mathcal{S}=\mathcal{S}_{1}\Cup\mathcal{S}_{2}$$
associated with the square root $j(\lambda)$. Each sheet of this
surface is determined by the sign of $j(\lambda)$.
\begin{eqnarray}\label{spec}
\mathcal{S}_{1}\colon \im j(\lambda)>0, \qquad
\mathcal{S}_{2}\colon \im j(\lambda)<0.
\end{eqnarray}
Half of the columns of the Jost solutions are analytic functions
of $\lambda$ on the first sheet and the other on the second sheet.
\begin{equation}\label{}
\psi(x,\lambda)=(|\psi^{-}(x,\lambda)\rangle,
|\psi^{+}(x,\lambda)\rangle),\qquad \phi(x,\lambda)=
(|\phi^{+}(x,\lambda)\rangle,   |\phi^{-}(x,\lambda)\rangle),
\end{equation}
where $|\psi^{\pm}\rangle$ and $|\psi^{\pm}\rangle$ denote a $r
\times 2r$ matrix composed of the corresponding $r$ columns of the
Jost solutions. The superscript ''$^{+}$'' means analyticity on
the first sheet and ''$^{-}$'' - analyticity on the second sheet.
Next, we can construct FAS on each of the sheets by simply
combining the blocks of the Jost solutions with the same
analyticity properties.
\begin{equation}\label{}
\chi^{+}(x,\lambda)\equiv\left(|\phi^{+}\rangle,|\psi^{+}
\rangle\right)(x,\lambda),\qquad
\chi^{-}(x,\lambda)\equiv\left(|\psi^{-}\rangle, |\phi^{-}\rangle
\right)(x,\lambda)
\end{equation}

Let us write down the FAS $\chi^{+}(x,\lambda)$, analytic on the
sheet $\mathcal{S}_{1}$ and $\chi^{-}(x,\lambda)$, analytic on the
sheet $\mathcal{S}_{2}$ using appropriate decompositions of the
scattering matrix (\ref{Scatt}), which consists of the same upper
(lower) block-triangular functions $\textbf{S}^{\pm}$ and
$\textbf{T}^{\pm}$ as they are in the v.b.c. case
 \cite{VSG_basics}:
\begin{eqnarray}\label{chi+m_equal}
\chi^{\pm}(x,\lambda)=\psi(x,\lambda)\textbf{T}^{\mp}
=\phi(x,\lambda)\textbf{S}^{\pm}
\end{eqnarray}
These triangular factors are:
\begin{eqnarray}
 \textbf{S}^{+}=\left(
\begin{array}{cc}
  \openone & \textbf{d}^{-} \\
  0 & \textbf{c}^{+} \\
\end{array}
\right), \quad \textbf{T}^{-}=\left(%
\begin{array}{cc}
  \textbf{a}^{+} & 0 \\
  \textbf{b}^{+} & \openone \\
\end{array}
\right), \qquad
  \textbf{S}^{-}=\left(
\begin{array}{cc}
  \textbf{c}^{-} & 0 \\
  -\textbf{d}^{+} & \openone \\
\end{array}
\right),\qquad \textbf{T}^{+}=\left(%
\begin{array}{cc}
  \openone & -\textbf{b}^{-} \\
  0 & \textbf{a}^{-} \\
\end{array}
\right)
\end{eqnarray}
and  can be viewed also as generalized Gauss decompositions of the
$T(\lambda)$.
\begin{equation}
T(\lambda)=\textbf{T}^{-}(\lambda)\widehat{\textbf{S}}^{+}(\lambda)=
\textbf{T}^{+}(\lambda)\widehat{\textbf{S}}^{-}(\lambda).
\end{equation}
Here and after the hat $^{\;\widehat{}\;}$ means taking the
inverse matrix. We can use for the scattering matrix the same
block-matrix structure as in v.b.c. case \cite{LOMI}:
\begin{eqnarray}\label{Scatt}
\phi(x,\lambda)=\psi(x,\lambda)T(\lambda),\quad T(\lambda)=\left(%
\begin{array}{cc}
  \textbf{a}^{+}(\lambda) & -\textbf{b}^{-}(\lambda) \\
  \textbf{b}^{+}(\lambda) & \textbf{a}^{-}(\lambda)\\
\end{array}%
\right) &&
%\psi(x,\lambda)=\phi(x,\lambda)\widehat{T}(\lambda),
\quad\widehat{T}(\lambda)=\left(%
\begin{array}{cc}
  \textbf{c}^{-}(\lambda) & \textbf{d}^{-}(\lambda) \\
  -\textbf{d}^{+}(\lambda) & \textbf{c}^{+}(\lambda)\\
\end{array}
\right)
\end{eqnarray}
The elements of the inverse matrix are defined as follows:
\begin{eqnarray*}
\textbf{c}^{-}(\lambda)&=&\widehat{\textbf{a}}^{\,+}(\lambda)
(\openone+\rho^{-}\rho^{+})^{-1}=
(\openone+\tau^{+}\tau^{-})^{-1}\widehat{\textbf{a}}^{\,+}(\lambda)\\
\textbf{d}^{-}(\lambda)&=&\widehat{\textbf{a}}^{\,+}(\lambda)
\rho^{-}(\lambda)\,(\openone+\rho^{+}\rho^{-})^{-1}=
(\openone+\tau^{+}\tau^{-})^{-1}\tau^{+}(\lambda)
\widehat{\textbf{a}}^{\,-}(\lambda)\notag\\
\textbf{c}^{+}(\lambda)&=&\widehat{\textbf{a}}^{\,-}(\lambda)
(\openone+\rho^{+}\rho^{-})^{-1}=
(\openone+\tau^{+}\tau^{-})^{-1}\widehat{\textbf{a}}^{\,-}(\lambda)\\
\textbf{d}^{+}(\lambda)&=&\widehat{\textbf{a}}^{\,-}(\lambda)
\rho^{+}(\lambda)\,(\openone+\rho^{-}\rho^{+})^{-1}=
(\openone+\tau^{-}\tau^{+})^{-1}\tau^{-}(\lambda)
\widehat{\textbf{a}}^{\,+}(\lambda)
\end{eqnarray*}
Here
$\rho^{\pm}(\lambda)=\textbf{b}^{\pm}(\lambda)\widehat{\textbf{a}}
^{\,\pm}(\lambda)=\widehat{\textbf{c}}^{\pm}(\lambda)\textbf{d}^{\,\pm}
(\lambda)$ and
$\tau^{\pm}(\lambda)=\widehat{\textbf{a}}^{\,\pm}(\lambda)
\textbf{b}^{\,\mp}(\lambda)=\textbf{d}^{\mp}(\lambda)
\widehat{\textbf{c}}^{\,\pm}(\lambda)$ are the multicomponent
generalizations of the reflection $\rho^{\pm}$, $\tau^{\pm}$
coefficients. (for the scalar case see
 \cite{ZSh1,ZSh2,ZaMa*74bR}).

Given the potential $Q(x)$ one can obtain the Jost solutions
uniquely. The Jost solutions in turn determine uniquely the
scattering matrix $T(\lambda)$ and its inverse
$\widehat{T}(\lambda)$. $Q(x)$ contains at most
$|\triangle^{1}_{+}|$ independent complex-valued functions of $x$.
Thus it is natural to expect that at most $|\triangle^{+}_{1}|$ of
the coefficients of $T(\lambda)$ for $\lambda\in\bbbr_{m}$,
instead of $(2r)^2$, will be independent. Here
$|\triangle^{+}_{1}|$ is the number of roots in
$\triangle^{+}_{1}$,i.e. $
    |\triangle^{+}_{1}|=r(r+1)/2
$ for ${\bf C}_{r}$ and $
    |\triangle^{+}_{1}|=r(r-1)/2
$
for ${\bf D}_{r}$. The continuous spectrum $\bbbr_{m}=(-\infty,
-m)\cup (m, \infty)$ is determined by the condition $|\lambda|\geq
m$.

The set of independent coefficients of $T(\lambda)$ are known as
the set of minimal scattering data $\mathcal{T}$. They were
introduced by Kaup for the Z-Sh system associated with
$\mathfrak{g}\simeq sl(2)$ and v.b.c.. He proved that
$a^{\pm}(\lambda)$ can be recovered from $\mathcal{T}$ using the
analyticity properties, i.e. the so-called dispersion relation.
The same problem for the generalized Z-Sh system with c.b.c. is
more difficult. Here we just introduce
$\mathcal{T}_{i}=\mathcal{T}_{i,c}\cup\mathcal{T}_{i,d}$ as the
proper generalization of the minimal set of scattering data:
\begin{eqnarray*}
 \mathcal{T}_{1,c}\equiv\{ \rho^{+}_{\alpha}(\lambda),
\rho^{-}_{\alpha}(\lambda),\quad \lambda\in\bbbr_{m}  \}, &&
\mathcal{T}_{1,d}\equiv\{\rho^{\pm}_{\alpha}(\lambda_{j}^{\pm}),
   \lambda_{j}^{\pm}\}_{j=1}^{N}\\ \mathcal{T}_{1,c}\equiv\{
  \tau^{+}_{\alpha}(\lambda), \tau^{-}_{\alpha}(\lambda),\quad
  \lambda\in\bbbr_{m}  \}, &&
 \mathcal{T}_{2,d}\equiv\{\tau^{\pm}_{\alpha}(\lambda_{j}^{\pm}),
  \lambda_{j}^{\pm}\}_{j=1}^{N}
\end{eqnarray*}
where $\alpha\in\triangle^{+}_{1}$.  The reconstruction of the
diagonal blocks $\textbf{a}^{\pm}(\lambda)$ from their analyticity
properties requires a solution of $r\times r$ matrix-valued
Riemman-Hilbert problem. Here $\lambda_{j}^{\pm}$ are discrete
eigenvalues of $L$. The sets $\mathcal{T}_{i,c}$ characterizing
the continuous spectrum need to be completed by the sets
$\mathcal{T}_{i,d}$ characterizing the discrete spectrum of $L$
which in turn requires the knowledge of the dressing factors.
These problems will be addressed elsewhere.

\section{Wronskian relations}
Let the class of allowed potentials $\mathcal{M}$ be a slice of
$O_{\sigma_3 }$ determined by additional constraints: i.) any
generic element $F(x)=P_{\sigma_3 }F(x)$ of $\mathcal{M}$ is
matrix-valued function which vanishes fast enough for
$|x|\to\infty$ and ii.) the phase factor $V$ which connect the
asymptotic values of the potential $Q_{+}=V^{\dagger}Q_{-}V$ is an
integral of motion. The derivative of the potential $Q_{x}(x,t)$
belongs to the class of allowed potentials. The variation of the
potential $\delta Q(x,t)$ is an allowed potential provided it
satisfies the second additional condition. The mapping
$\mathcal{F}:\mathcal{M}\to \mathcal{L} $ between the class of
allowed potentials $\mathcal{M}$ and the scattering data
$\mathcal{L}$ of $L$ is analyzed by means of Wronskian relations
\cite{Calogero1,Calogero2}. These relations allow us to formulate
the main result of this work, i.e. that the ISM is a generalized
Fourier transform in the case of {\bf C.I} and {\bf D.III}-type
symmetric spaces. They also serve to introduce the skew-scalar
product
\begin{equation}\label{skew-scal}
    \Biglb A(x),B(x)\Bigrb=\frac{1}{2}\int
dx\,\left\langle\,A(x)\,,\,[\,\sigma_{3}\,,B(x)\,]\, \right\rangle
\end{equation} which is non-degenerate for $A(x), B(x) \in
\mathcal{M}$ and provides it with symplectic structure. We start
with the identity:
\begin{equation}\label{W1}
\left\langle\widehat{\chi}\,\left(Q(x,t)-\lambda\sigma_3 \right)
\,\chi(x,\lambda)\,,\,E_{\pm\alpha}\,
\right\rangle|_{x=-\infty}^{\infty}=-i\int_{-\infty}^{\infty}dx
\;\left\langle
\,\frac{i}2\left[\,\sigma_{3}\,,\sigma_{3}\,Q_{x}\right]\,,\,P_{\sigma_{3}}
\,\chi\,E_{\pm\alpha}\,\widehat{\chi}(x,\lambda)\,\right\rangle
\end{equation}
where $\chi(x,\lambda) $ can be any fundamental solution of $L$.
For convenience we choose them to be the FAS introduced above. The
l.h.side of (\ref{W1}) can be calculated explicitly by using the
asymptotics of FAS for $x \to \pm\infty$. It would be expressed by
the matrix elements of the scattering matrix $T(\lambda)$, i.e. by
the scattering data of $L$ as follows:
\begin{eqnarray}\label{koef_Qx}
\biglb P_{\sigma_{3}}\,\chi^{+}(x,\lambda)\,E_{\alpha}\,\widehat{\chi}^{\,+},
  \sigma_{3} \,Q_{x}\bigrb= -j(\lambda)\left\langle
\widehat{\textbf{T}}^{-}\sigma_3 \textbf{T}^{-} \,,\,E_{\alpha}
\right\rangle=
2j(\lambda)\textbf{b}^{+}_{\alpha},\qquad \alpha\in\triangle^{+}_{1}&&\notag\\
\biglb P_{\sigma_{3}}\,\chi^{+}(x,\lambda)\,E_{-\alpha}\,\widehat{\chi}^{\,+},\sigma_{3}
\,Q_{x}\bigrb=  j(\lambda)\left\langle
\widehat{\textbf{S}}^{+}\sigma_3 \textbf{S}^{+} \,,\,E_{-\alpha}
\right\rangle=
2j(\lambda)\textbf{d}^{-}_{-\alpha},\qquad \alpha\in\triangle^{+}_{1}&&\notag\\
\biglb
P_{\sigma_{3}}\,\chi^{-}(x,\lambda)\,E_{\alpha}\,\widehat{\chi}^{\,-},\sigma_{3}
\,Q_{x}\bigrb=  j(\lambda)\left\langle
\widehat{\textbf{S}}^{-}\sigma_3 \textbf{S}^{-} \,,\,E_{\alpha}
\right\rangle=
2j(\lambda)\textbf{d}^{+}_{\alpha},\qquad \alpha\in\triangle^{+}_{1}&&\notag\\
\biglb
P_{\sigma_{3}}\,\chi^{-}(x,\lambda)\,E_{-\alpha}\,\widehat{\chi}^{\,-},\sigma_{3}
\,Q_{x}\bigrb=-j(\lambda)\left\langle
\widehat{\textbf{T}}^{+}\sigma_3 \textbf{T}^{+} \,,\,E_{-\alpha}
\right\rangle= 2j(\lambda)\textbf{b}^{-}_{-\alpha},\qquad
\alpha\in\triangle^{+}_{1}&&
\end{eqnarray}

The second set of Wronskian relations which we consider relate the
variation of the potential $\delta Q $ to the corresponding
variations of the scattering data $\delta\rho$ and $\delta\tau$.
For this purpose we use the identity:
\begin{equation}\label{W2}
    \left\langle\widehat{\chi}\,\delta\chi(x,\lambda)\,,\,E_{\pm\alpha}\,
\right\rangle |_{x=-\infty}^{\infty}=\int_{-\infty}^{\infty}dx
\;\left\langle \,\frac{i}2\left[\,\sigma_{3}\,,\sigma_{3}\,\delta
Q\right]\,,\,P_{\sigma_{3}}\,\chi(x,\lambda)
\,E_{\pm\alpha}\,\widehat{\chi}\,\right\rangle
\end{equation}
If we assume that the variation of the phase factor $\delta\,V$
vanishes we arrive at:
\begin{eqnarray}\label{koef_deltaQ}
\biglb
P_{\sigma_{3}}\,\chi^{+}(x,\lambda)\,E_{\alpha}\,\widehat{\chi}^{\,+},\sigma_{3}\,\delta
Q\,\bigrb= -i\left\langle
\widehat{\textbf{T}}^{-}\delta\textbf{T}^{-} \,,\,E_{\alpha}
\right\rangle=
i(\delta\textbf{$\rho$}^{+}\textbf{a}^{+})_{\alpha},\qquad
\alpha\in\triangle^{+}_{1}&&\notag\\
\biglb
P_{\sigma_{3}}\,\chi^{+}(x,\lambda)\,E_{-\alpha}\,\widehat{\chi}^{\,+},\sigma_{3}
\,\delta
Q\,\bigrb=
 i\left\langle
\widehat{\textbf{S}}^{+}\delta \textbf{S}^{+} \,,\,E_{-\alpha}
\right\rangle=
i(\delta\textbf{$\tau$}^{+}\textbf{c}^{+})_{-\alpha} ,\qquad
\alpha\in\triangle^{+}_{1}&&\notag\\
\biglb
P_{\sigma_{3}}\,\chi^{-}(x,\lambda)\,E_{\alpha}\,\widehat{\chi}^{\,-},\sigma_{3}\,\delta
Q\,\bigrb=
 i\left\langle
\widehat{\textbf{S}}^{-}\delta \textbf{S}^{-} \,,\,E_{\alpha}
\right\rangle=
i(\delta\textbf{$\tau$}^{-}\textbf{c}^{-})_{\alpha},\qquad
\alpha\in\triangle^{+}_{1}&&\notag\\
\biglb
P_{\sigma_{3}}\,\chi^{-}(x,\lambda)\,E_{-\alpha}\,\widehat{\chi}^{\,-},\sigma_{3}
\,\delta
Q\,\bigrb=
 -i\left\langle
\widehat{\textbf{T}}^{+}\delta\textbf{T}^{+} \,,\,E_{-\alpha}
\right\rangle=
i(\delta\textbf{$\rho$}^{-}\textbf{a}^{-})_{-\alpha} ,\qquad
\alpha\in\triangle^{+}_{1}&&
\end{eqnarray}

These relations are basic for the analysis of the related NLEE and
their Hamiltonian structures. The above identities also allow us
to introduce the proper generalizations of the usual Fourier
exponential functions. Let us introduce the set of "squared
solutions":
\begin{eqnarray}\label{Squared_Solutions}
 {\Phi}_{\alpha}^{\pm}(x,\lambda)=P_{\sigma_{3}}\,\chi^{\pm}(x,\lambda)
E_{\pm\alpha}\widehat{\chi}^{\pm}(x,\lambda)\,,&&\text{for}\;
\alpha\in\triangle_{1}^{+}\\
 {\Psi}_{\alpha}^{\pm}(x,\lambda)=P_{\sigma_{3}}\,\chi^{\pm}(x,\lambda)
E_{\mp\alpha}\widehat{\chi}^{\pm}(x,\lambda)\,,&&\text{for}\;
\alpha\in\triangle_{1}^{+}\\
&&\notag\\
 {\Theta}_{\alpha}^{\pm}(x,\lambda)=P_{\sigma_{3}}\,\chi^{\pm}(x,\lambda)
E_{\pm\alpha}\widehat{\chi}^{\pm}(x,\lambda)\,,&&\text{for}\;
\alpha\in\triangle_{0}^{+}\\
 {\Xi}_{\alpha}^{\pm}(x,\lambda)=P_{\sigma_{3}}\,\chi^{\pm}(x,\lambda)
E_{\mp\alpha}\widehat{\chi}^{\pm}(x,\lambda)\,,&&\text{for}\;
\alpha\in\triangle_{0}^{+}\\
 {\Upsilon}_{k}^{\pm}(x,\lambda)=P_{\sigma_{3}}\,\chi^{\pm}(x,\lambda)
H_{k}\widehat{\chi}^{\pm}(x,\lambda)\,,&&\text{for}\;
k=1,...,r\end{eqnarray}

These are the "squared solutions" of the Lax operator $L$
connected with simple Lie algebra $\mathfrak{g}$. They are
constructed by means of FAS $\chi^{\pm}(x, \lambda)$ and the
Cartan-Weyl basis of the algebra and are analytic functions of
$\lambda$ on the corresponding sheets of the spectral surface. The
equations that ${\Phi}_{\alpha}^{\pm}$ an ${\Psi}_{\alpha}^{\pm}$
satisfy are a direct consequence of the fact that FAS and their
inverse satisfy the Z-Sh system system:
\begin{equation}\label{SSeqn}
    i\frac{d\,{\Phi}_{\alpha}^{\pm}}{dx}+\left[Q(x)-\lambda\,\sigma_{3},
    {\Phi}_{\alpha}^{\pm}(x,\lambda)\right]=0,\qquad
    i\frac{d\,{\Psi}_{\alpha}^{\pm}}{dx}+\left[Q(x)
    -\lambda\,\sigma_{3}, {\Psi}_{\alpha}^{\pm}(x,\lambda)\right]=0
\end{equation}

The "squared solutions" also serve as building blocks of the Green
function for $L$ \cite{VSG86,VSG_Thesis,VSG_basics}:

\begin{equation}\label{green}
\mathbf{G}^{\pm}(x,y,\lambda)=G^{\pm}_{1}(x,y,\lambda)\theta(y-x)
-G^{\pm}_{2}(x,y,\lambda)\theta(x-y),
\end{equation}
where
\begin{eqnarray}\label{green1}
G^{\pm}_{1}(x,y,\lambda)&=&\sum_{\alpha\in\triangle^{+}_{1}}
{\Phi}_{\alpha}^{\pm}(x,\lambda)\otimes
{\Psi}_{\alpha}^{\pm}(y,\lambda)\\
\label{green2}
G^{\pm}_{2}(x,y,\lambda)&=&\sum_{\alpha\in\triangle^{+}_{1}}
{\Psi}_{\alpha}^{\pm}(x,\lambda)\otimes
{\Phi}_{\alpha}^{\pm}(y,\lambda)
+\sum_{\alpha\in\triangle^{+}_{0}\cup\triangle^{+}_{1}}
{\Xi}_{\alpha}^{\pm}(x,\lambda) \otimes
{\Theta}_{\alpha}^{\pm}(y,\lambda)\notag\\
&+&\sum_{k=1}^{r}\Upsilon_{k}^{\pm}(x,\lambda)\otimes
 \Upsilon_{k}^{\pm}(y,\lambda)
\end{eqnarray}

\begin{figure}[htb]
\centerline{\includegraphics[width=8cm,height=8cm,clip]{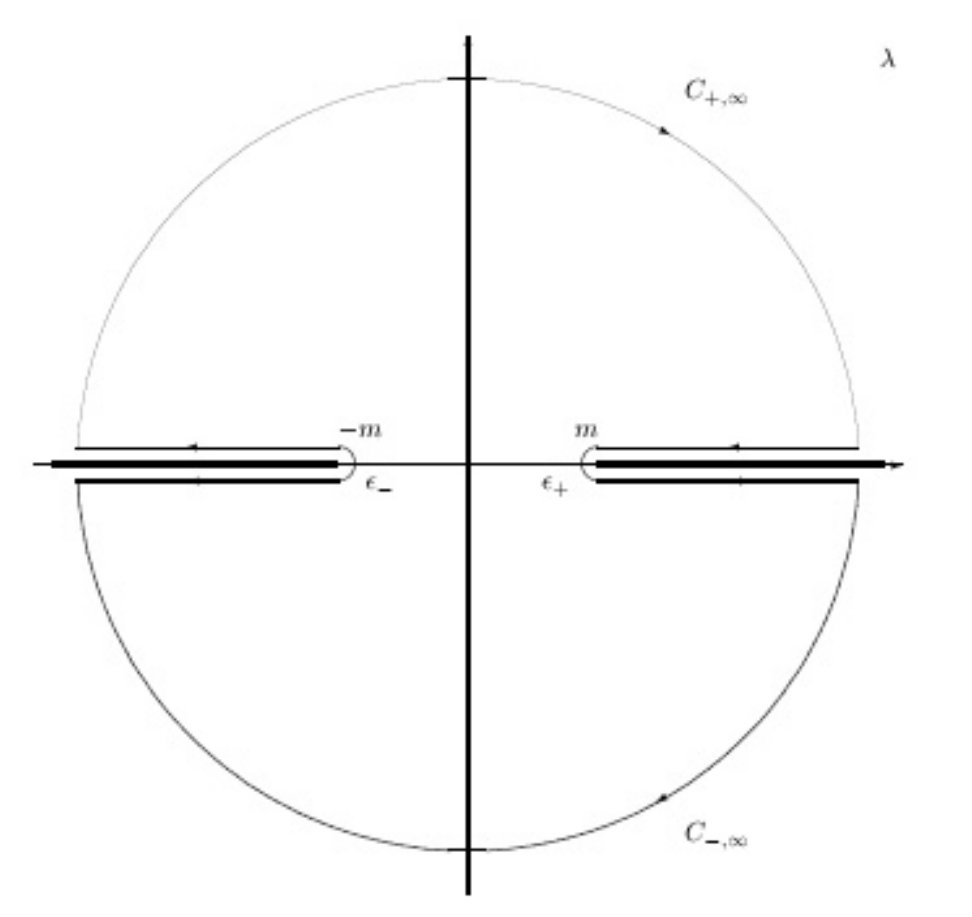}}
\caption{The contour along which we integrate lies completely on
the first sheet of the 2-sheeted spectral surface associated with
the square root $j(\lambda)=\sqrt{\lambda^2-m^2}$} \label{fig3}
\end{figure}
and $\theta(x)$ is the usual step function.

\section{Completeness relation and evolution}

The main result in this section is that the sets
$\left\{{\Phi}_{\alpha}^{\pm}\right\}$ and $\left\{
{\Psi}_{\alpha}^{\pm}\right\}$ form complete sets of functions in
$\mathcal{M}$. The idea of the proof is simple. Apply the contour
integration method along a proper contour (see figure.\ref{fig3})
to a conveniently chosen Green function (\ref{green}). From the
Cauchy theorem we have
\begin{equation}\label{contour_Int}
\frac{1}{2\pi i}\oint_{C}
\mathbf{G}^{+}(x,y,\lambda)d\lambda=\sum_{k=1}^{N}\res_{\lambda=
\lambda^{\pm}_{k}}\,\mathbf{G}^{+}(x,y,\lambda)
\end{equation}
Integrating along the contours we treat separately the
contribution from the infinite semi-arcs and the ones from the
continuous spectrum ${R}_{m}=C_1 \cup C_2$ which is composed of
the cuts $C_1=(-\infty, -m)$ and $C_2=(m, \infty)$. Special care
must be taken for the end points $\lambda=\pm m$ of the spectrum.
Assuming that the end points of the spectrum give no contribution
\begin{equation}\label{obshtopolojenie}
\lim_{\varepsilon\to
0}\left(\int_{\varepsilon_{+}}+\int_{\varepsilon_{-}}\right)\,\mathbf{G}^{+}
d\lambda=0
\end{equation}
where with index $\varepsilon_{\pm}$ we have denoted the integrals
along the infinitesimal semi-arcs around the end points of the
spectrum, we obtain the following completeness relation
\begin{eqnarray}\label{completeness}
\delta(x-y)\Pi_{\sigma_{3}}&=&\frac{1}{\pi
}\sum_{\alpha\in\triangle^{+}_{1}}\,\int_{
{R}_{m}}d\lambda\,\left\{ {\Phi}_{\alpha}^{+}(x,\lambda)\otimes
 {\Psi}_{\alpha}^{+}(y,\lambda)- {\Phi}_{\alpha}^{-}(x,\lambda)\otimes
 {\Psi}_{\alpha}^{-}(y,\lambda)\right\}\notag\\
&-&2i\sum_{\alpha\in\triangle^{+}_{1}}\,\sum_{k=1}^{N}\;\left\{\frac{d}{d\lambda}
{\Phi}_{\alpha}^{+}(x,\lambda)\right|_{\lambda=\lambda^{+}_{k}}\left.
\otimes {\Psi}_{\alpha}^{+}(y,\lambda)+
{\Phi}_{\alpha}^{+}(x,\lambda) \otimes\frac{d}{d\lambda}
{\Psi}_{\alpha}^{+}(y,\lambda)\right|_{\lambda=\lambda^{+}_{k}}\left.\right\}\notag
\end{eqnarray}
Here
$\Pi_{\sigma_{3}}=\sum_{\alpha\in\triangle^{+}_{1}}[E_{\alpha}\otimes
E_{-\alpha}-E_{-\alpha}\otimes E_{\alpha}]$. The assumption that
we have made is that $\lambda_{j}^{+}$ are simple poles of the
"squared solutions" $\;{\Phi}_{\alpha}^{+}$ and $
{\Psi}_{\alpha}^{+}$.

Using the completeness relation one can expand any generic element
of the phase space $\mathcal{M}$ over each of the complete sets of
"squared solutions" $\;{\Psi}_{\alpha}^{\pm}$ and
${\Phi}_{\alpha}^{\pm}$. This relation is utilized with the help
of the following the trick
\begin{equation}\label{trick}
-\frac12\tr_{1}\left\{([\sigma_{3},F(x)]\otimes
\openone)\Pi_{\sigma_{3}}
\right\}=\frac12\tr_{2}\left\{\Pi_{\sigma_{3}}(\openone\otimes
[\sigma_{3},F(x)]) \right\}=F(x)
\end{equation}
where $\tr_{1}$ (and $\tr_{2}$) mean taking the trace of the
elements in the first (or in the second) position of the tensor
product.

The completeness relation (\ref{completeness}) allows to establish
one-to-one correspondence between the elements of $\mathcal{M}$,
such as $Q_{x}$ and $Q_{t}$, and its expansion coefficients. It is
also directly related to the spectral decompositions of the
generating (recursion) operators $\Lambda_{\pm}$. These operators
are the ones whose eigenfunctions are the "squared solutions".
Their derivation starts by introducing the splitting of the object
$e^{\pm}_{\alpha}=\chi^{\pm}(x,\lambda)
E_{\pm\alpha}\widehat{\chi}^{\pm}(x,\lambda)$ into block diagonal
and block off-diagonal parts:
\begin{equation}\label{splitting}
    e^{\pm}_{\alpha}(x, \lambda)=e^{\,d,\,\pm}_{\alpha}(x,
    \lambda)+\Phi^{\pm}_{\alpha}(x,
    \lambda),\qquad e^{\,d,\,\pm}_{\alpha}(x,
    \lambda)=\left(\openone-P_{\sigma_{3}}
    \right)e^{\pm}_{\alpha}(x, \lambda)
\end{equation}
end use the equation it satisfies
\begin{eqnarray}\label{eeqn}
    i\frac{d\,e^{\pm}_{\alpha}}{dx}+\left[Q(x)-\lambda\,\sigma_{3},
    e^{\pm}_{\alpha}(x, \lambda)\right]=0
\end{eqnarray}
Thus equation (\ref{eeqn}) splits into
\begin{equation}\label{purvoto}
  i\frac{d\,e^{\,d,\,\pm}_{\alpha}}{dx}+\left[Q(x), \Phi^{\pm}_{\alpha}(x, \lambda)\right]=0
\end{equation}
\begin{equation}\label{vtoroto}
  i\frac{d\,\Phi^{\pm}_{\alpha}}{dx}+\left[Q(x), e^{\,d,\,\pm}_{\alpha}(x, \lambda)\right]=
  \lambda\left[\sigma_{3}, \Phi^{\pm}_{\alpha}(x, \lambda)\right]
 \end{equation}
Equation (\ref{purvoto}) can be integrated formally with the
result
\begin{equation}\label{result}
    e^{\,d,\,\pm}_{\alpha}(x,
    \lambda)=C_{\alpha;\,\epsilon}^{d,\,\pm}(\lambda)+i\int_{\epsilon\infty}^{x}
    dy\left[Q(y), \Phi^{\pm}_{\alpha}(y, \lambda)\right],
\end{equation}
\begin{equation}\label{C}
    C_{\alpha;\,\epsilon}^{d,\,\pm}(\lambda)=\lim_{x\to\epsilon\infty}e^{\,d,\,\pm}_{\alpha}(x,
    \lambda),\qquad \epsilon=\pm1
\end{equation}
Next insert (\ref{result}) into (\ref{vtoroto}) and act on both
sides by $\ad^{-1}$. This gives us:
\begin{equation}\label{}
    \left(\Lambda_{\pm}-\lambda\right)\Phi_{\alpha}^{\pm}(x,\lambda)=i\left[C_{\alpha;\,\epsilon}^
    {d,\,\pm}(\lambda),\ad^{-1}Q(x)\right],
\end{equation}
where the generating operators $\Lambda_{\pm}$ are given by:
\begin{equation}\label{}
    \Lambda_{\pm} \Xi(x)=\ad^{-1}\left\{ i\frac{d
 \Xi}{dx}+i\left[Q(x)\,,\,\int_{\pm\infty}^{x}dy\,\left[
 Q(y)\,,\,\Xi(y) \right] \right] \right\}
\end{equation}
Thus $\Psi_{\alpha}^{\pm}$ (resp. $\Phi_{\alpha}^{\pm}$ ) will be
eigenfunctions of $\Lambda_{+}$ (resp. $\Lambda_{-}$) if and only
if $C_{\alpha;\,\epsilon}^{d,\,\pm}(\lambda)=0$. Evaluating the
limit of (\ref{C}) for all $\alpha$ in the specific case
(\ref{obshtopolojenie}) we find:

\begin{eqnarray}\label{LambdaSq}
\left( \Lambda_{+}-\lambda\right)\Psi_{\alpha}^{\pm}(x,\lambda)=0
  &&\qquad
  \left(\Lambda_{+}-\lambda_{j}^{\pm}\right)\Psi_{\alpha}^{\pm}(x,\lambda_{j}^{\pm})=0,\qquad \alpha\in\triangle^{+}_{1}\\
\left( \Lambda_{-}-\lambda\right)\Phi_{\alpha}^{\pm}(x,\lambda)=0
  &&\qquad
  \left(\Lambda_{-}-\lambda_{j}^{\pm}\right)\Phi_{\alpha}^{\pm}(x,\lambda_{j}^{\pm})=0,\qquad \alpha\in\triangle^{+}_{1}
\end{eqnarray}
This result can be generalized for arbitrary $f(\Lambda_{\pm})$:
\begin{eqnarray}\label{fLambdaSq}
  \left( f(\Lambda_{+})-f(\lambda)\right)\Psi_{\alpha}^{\pm}(x,\lambda)=0
  &&\qquad
  \left(f(\Lambda_{+})-f(\lambda_{j}^{\pm})\right)\Psi_{\alpha}^{\pm}(x,\lambda_{j}^{\pm})=0,\qquad \alpha\in\triangle^{+}_{1}\\
  \left( f(\Lambda_{-})-f(\lambda)\right)\Phi_{\alpha}^{\pm}(x,\lambda)=0
  &&\qquad
  \left(f(\Lambda_{-})-f(\lambda_{j}^{\pm})\right)\Phi_{\alpha}^{\pm}(x,\lambda_{j}^{\pm})=0,\qquad \alpha\in\triangle^{+}_{1}
\end{eqnarray}

The class of higher MNLS on symmetric spaces of {\bf C.I} and {\bf
D.III}-type and with c.b.c. can be put down in terms of the
derivative of the potential  $Q_{t}$  with respect to the
evolution parameter and the dispersion law $f(\lambda)=-2\lambda$
 \cite{FaTa,VSG_basics} as follows:
\begin{equation}\label{MNShConst}
    i\ad^{-1}\frac{\partial}{\partial t}Q+f(\Lambda)\ad^{-1}Q_{x}=0
\end{equation}
Substituting the objects in this formula with their expansions
over the "squared solutions" we obtain equations for the evolution
of the scattering data. The expansion coefficients of
$\ad^{-1}Q_{t}$ and $\ad^{-1}Q_{x}$ on the continuous spectrum
turn out to be exactly the minimal set of scattering data. The
evolution for the reflection and transition coefficients is
provided by
\begin{equation}\label{}
i\frac {\partial \textbf{$\rho$}^{\pm}}{\partial t}\pm
f(\lambda)\,j(\lambda)\textbf{$\rho$}^{\pm}(t,\lambda)=0,\qquad
i\frac {\partial \textbf{$\tau$}^{\pm}}{\partial t}\mp
f(\lambda)\,j(\lambda)\textbf{$\tau$}^{\pm}(t,\lambda)=0\qquad
\lambda\in\bbbr_{m}  .
\end{equation}

The observation that the scattering data evolves trivially is
visible from the equation depicting the evolution of the
scattering matrix $T(\lambda)$. This equation is a result of  the
compatibility condition (\ref{commutator}) and the fact that the
two Jost solutions $\psi$ and $\phi$ are solutions of the second
operator $Ì$ of the Lax pair in  the Z-Sh system (\ref{L}). Acting
with $i\frac{d}{dt}$ on $T(\lambda)$ (\ref{T}), we get:
\begin{equation}\label{uravnScattMatrix}
    i\frac{d}{dt}T(t,
    \lambda)-2\lambda j(\lambda)\left[\,\sigma_{3},T(\lambda)\right]=0,
\end{equation}
where $f(\lambda)=-2\lambda$ is the dispersion law for the MNLS
with c.b.c.. This equation for MNLS can also be derived from the
explicit form of the Lax representation (\ref{M}) by evaluating
the limit $\lim_{x\to\pm\infty}M\psi=0$. For the $r \times r$
blocks making up the scattering matrix we have:
\begin{equation}\label{}
    \frac {\partial \textbf{a}^{\pm}(t,\lambda)}{\partial t}=0,\qquad
    i\frac {\partial \textbf{b}^{\pm}}{\partial t}\mp 2\lambda\,j(\lambda)\textbf{b}^{\pm}(t,\lambda)=0
\end{equation}
These equations have obvious solutions. From the first equation is
clear that the diagonal blocks are conserved and their invariants
- upper (lower) principal minors as well as their determinants are
generating functionals of the special series of local infinitely
many integrals of motion $I_{k}$:
\begin{equation}
\ln \det \textbf{a}^{\pm}(\lambda)=\sum_{k=1}^{\infty}
\lambda^{-k}I_{k}
\end{equation}

This is the major idea of the ISM - a one-to-one change of
variables- from the multicomponent $\textbf{q}(x, t)$, in terms of
which the MNLS(\ref{MNLS}) is written, towards the scattering data
which satisfy linear evolution equations.

\section{Conclusion}

The result of this work is that the interpretation of the ISM as a
generalized Fourier transformation holds true in the case of Lax
operators with  constant boundary conditions on symmetric spaces connected
with the Lie algebras $\textbf{C}_{r}\simeq sp(2r)$ and
$\textbf{D}_{r}\simeq so(2r)$. The completeness relation of the
"squared solutions" of the generalized Z-Sh system in the case when the
Lax operator $L$ becomes self-adjoint is derived. The "squared solutions"
turn out to be generalizations of the usual Fourier exponential function
and eigenfunctions of the recursion operators $\Lambda_{\pm}$.  This
result allows one to prove that the corresponding NLEE results in linear
evolution for the scattering data. The recursion operators $\Lambda_{\pm}$
open the path towards the construction of {\it action-angle} variables for
the NLEE solvable with this generalization of the Z-Sh system and from
there the Hamiltonian formulation of these equations and their hierarchies
connected with $\Lambda_{\pm}$.

The physical applications of the NLS eq. both with vanishing and
non-vanishing boundary conditions is well known; the same holds true for
the Manakov system as well as for the $sp(4) $ MNLS with v.b.c., see
\cite{IMW}. It will be interesting to find physical applications also for
the MNLS with c.b.c.

\subsection*{Acknowledgements}
\addcontentsline{toc}{subsection}{Acknowledgements}

V.S.G wishes to acknowledge partial support from NSF of Bulgaria
under contract No.1410 and V.A.A. would like to express his
gratitude to the organizers of GAS05 for their hospitality and
financial support.

\subsection*{Appendix}
\addcontentsline{toc}{subsection}{Appendix}

The above definition of $\mathfrak{g}$ (\ref{algebra}) satisfies
the requirement that the Cartan subalgebra $\mathfrak{h}$ will be
made up of diagonal matrices. The Cartan generators $H_{k}$, dual
to $e_{k}$, are given by:
\begin{equation}\label{Kartan_gen}
H_{k}=E_{kk}-E_{\overline{k}\,\overline{k}}
\end{equation}
The element $\sigma_{3}=\sum_{k=1}^{r}H_{k}$, belongs to
$\mathfrak{h}$ and is dual to $\vec{a}$.

The root vectors in the typical representation are given by
\begin{eqnarray}\label{Weyl_gen2}
E_{e_{i}-e_{j}}=E_{ij}-(-1)^{i+j}E_{\overline{j}\,\overline{i}}\qquad
E_{e_{i}+e_{j}}=E_{i\,\overline{j}}-\epsilon_{0}(-1)^{i+j}E_{j\,\overline{i}}
\end{eqnarray}
where $1\leq i<j\leq r$ and $\epsilon_{0}=\pm 1$. Since
$\epsilon_{0}=1$ for ${\frak g}\simeq so(2r)$ equation
(\ref{Weyl_gen2}) gives vanishing result for $i=j$ which is
compatible with the fact that $2e_{i}$ are not roots of $so(2r)$;
for ${\frak g}\simeq sp(2r)$ $\epsilon_{0}=-1$ and equation
(\ref{Weyl_gen2}) by putting $i=j$ provides also an expression for
$E_{2e_{i}}$. However this expression is not normed with respect
to the Killing form $\langle E_{\alpha}, E_{-\alpha} \rangle=2$ .
The Weyl generators associated with the root $2e_{i}$ that we will
use are given by \cite{Hel}:
\begin{equation}\label{}
    E_{2e_{i}}=\sqrt{2}\,E_{i\,\overline{i}}
\end{equation}


\begin{thebibliography}{77}
\addcontentsline{toc}{section}{Literature}



\bibitem{AKNS} Ablowitz M., Kaup D., Newell A. and Segur H., {\it The Inverse Scattering
Transform - Fourier Analysis for Nonlinear Problems}, Stud. Appl.
Math {\bf 53} 249-315 (1974).


\bibitem{AbwSeg} Ablowitz M. and Seegur H., {\it Solitons and the Inverse Scattering Transform.
 SIAM Studies in Applied Mathematics} Philadelphia: SIAM, 1981


\bibitem{Calogero1} Calogero F.,Degasperis A., {\it Nonlinear Evolution Equations Solvable by
the Inverse Spectral Transform I} Nuovo Cim. {\bf 32B} 201-242
(1976).

\bibitem{Calogero2} Calogero F.,Degasperis A. {\it Nonlinear Evolution Equations Solvable by
the Inverse Spectral Transform II} Nuovo Cim. {\bf 39B} 1-54
(1976).


\bibitem{FK} Fordy A.P., Kulish P.P., {\it Nonlinear Schrodinger Equations and Simple
Lie Algebras} Commun.Math.Phys. {\bf 89} 427-443 (1983)


\bibitem{VSG86} Gerdjikov V.S.,{\it    Generalised Fourier transforms for  the soliton
     equations. Gauge covariant formulation.}
     Inverse Problems {\bf 2,} n.~1, 51-74, (1986).


\bibitem{VSG_Thesis} Gerdjikov V.S. , {\it Generating Operators for the Nonlinear Evolution Equations
of Soliton Type Related to the Semisimple Lie Algebras} Doctor of
Sciences Thesis, 1987, JINR, Dubna, USSR, (in Russian).

\bibitem{VSG_basics} Gerdjikov  V.S.. {\it Basic Aspects of Soliton Theory.}
"Geometry, Integrability and Quantization", Eds.: I.M.Mladenov,
A.C.Hirshfelt, SOFTEX Sofia, Bulgaria 2005.

\bibitem{VSG_BAN} Gerdjikov  V.S., {\it Selected Aspects of Soliton
Theory. Constant boundary conditions} (in press).

\bibitem{GGK} Gerdjikov V.S., Grahovski G.G., Kostov N.A. {\it Reductions of $N$-wave
interactions related to low-rank simple Lie algebras: I.
$\mathbb{Z}_2$-reductions} J. Phys. {\bf A34} 9425-9461 (2001)

\bibitem{LOMI} Gerdjikov V.S. , Kulish P.P. ,
{\it On the multicimponent nonlinear Schr\"odinger equation in the
case of non-vanishing boundary conditions}, Sci. Notes of LOMI
seminars {\bf 131}, 34--46, (1983).


\bibitem{Hel} Helgasson S. , {\it Differential Geometry, Lie Groups and Symmetric
Spaces}, Academic Press, New York, 1978

\bibitem{IMW} Ieda J., Miyakawa T., Wadati M., {\it Exact analysis of soliton dynamics in spinor
Bose-Einstein condensates} Phys. Rev. Lett. {\bf 93}, 194102
(2004).

\bibitem{DJ}Kaup D.J. {\it Closure of the Squared Zakharov-Shabat
Eigenstates}J. Math. Annal. Appl. {\bf 54}, n.3, 849-864, (1976).

\bibitem{KonVek}  Konotop V.V., Vekslerchik V.E., {\it Direct perturbation theory for dark
solitons} Phys.Rev. {\bf 49E}, 2397-2407 (1994)


\bibitem{Ma*76aR}
    Manakov S.V. , {\it On The Theory of Two-dimentional Stationary Self-focusing
    of Electromagnetic Waves}
   Sov. Phys. JETF {\bf 38}, 248-253, (1974).

\bibitem{Mik} Mikhailov  A.V., {\it The Reduction Problem and the Inverse Scattering
Problem} Physica D, 3{\bf D}, 73-117 (1981)



\bibitem{FaTa} Takhtadjan  L. A. and
Faddeev L. D., {\it Hamiltonian Approach to Soliton Theory}
(Springer-Verlag, Berlin, 1986).



\bibitem{ZaMa*74bR}   Zakharov V. E.,  Manakov S.V.,
   TMP {\bf 19}, No. 3, 332-343, 1974.               \\
   {\it On the complete integrability of the nonlinear
   Schr\"odinger equation.} (In Russian)


\bibitem{ZMNP} Zakharov V. E., Manakov  S. V.,  Novikov S. P., and
Pitaevskii L. I., {\it The Theory of Solitons. The Inverse
Transform Method }(Nauka, Moscow, 1980) (in Russian).

\bibitem{ZSh1} Zakharov V. E., Shabat  A.B., {\it Exact Theory of Two-dimentional Self-focusing
and One-dimentional Modulation of Waves in Nonlinear Media.},Sov.
Phys, JETF {\bf 34}, 62-69 (1972) (In Russian).


\bibitem{ZSh2} Zakharov V. E., Shabat A.B., {\it On the interaction of solitons
in stable medium.},Sov. Phys.  JETF {\bf 37}, 823-828 (1973) (In
Russian).


\end{thebibliography}
\end{document}